\documentclass{aastex631}
\usepackage{amsmath}

\usepackage{setspace}
\usepackage[caption=false]{subfig}

\received{March 19, 2025}
\revised{May 20, 2025}
\accepted{June 2, 2025}

\begin{document}

\title{Measuring Scaling Relations: Fitting Technique Matters}

\correspondingauthor{Bryanne McDonough}
\email{br.mcdonough@northeastern.edu}

\author[0000-0001-6928-4345]{Bryanne McDonough}
\affiliation{Department of Physics, Northeastern University, 360 Huntington Ave, Boston, MA, USA}
\affiliation{Institute for Astrophysical Research, Boston University, 725 Commonwealth Ave, Boston, MA, USA}

\author[0000-0002-0212-4563]{Olivia Curtis}
\affiliation{Department of Astronomy \& Astrophysics, The Pennsylvania State University, 251 Pollock Road, University Park, PA, USA}
\affiliation{Institute for Astrophysical Research, Boston University, 725 Commonwealth Ave, Boston, MA, USA}

\author[0000-0001-7917-7623]{Tereasa G. Brainerd}
\affiliation{Institute for Astrophysical Research, Boston University, 725 Commonwealth Ave, Boston, MA, USA}



\begin{abstract}
Scaling relationships, both integrated and spatially resolved, arise due to the physical processes that govern galaxy evolution and are frequently measured in both observed and simulated data. However, the accuracy and comparability of these measurements are hindered by various differences between studies such as spatial resolution, sample selection criteria, and fitting technique. 
Here, we compare variations of standard least squares techniques to the ridge line method for identifying spatially resolved scaling relations ($\Sigma_*-\Sigma_{\rm SFR}$, $\Sigma_*-\Sigma_{\rm gas}$, and $\Sigma_{\rm gas}-\Sigma_{\rm SFR}$) for TNG100 galaxies. 
We find that using the ridge line technique to fit these scaling relations with a double linear function results in significantly better fits than fitting with ordinary least squares. 
We further illustrate the utility of the ridge line technique with an investigation into the dependence of rSFMS measurements on spatial resolution and smoothing scale. 
Specifically, we find that the slope of the rSFMS at low-$\Sigma_*$ is independent (within $2\sigma$) of spatial resolution and smoothing scale. Finally, we discuss the need for a consistent re-analysis of resolved scaling relations in the literature and physically motivate adoption of the ridge line technique over other fitting methods.


\end{abstract}




\section{Introduction} \label{sec:intro}

Scaling relationships are strong correlations between galaxy properties that arise due to the physical mechanisms that drive galaxy evolution. As ever higher spatial resolutions have become possible, many relations previously identified between integrated (i.e., global) galaxy properties have been found to have spatially resolved analogs, down to $\sim100$s of parsecs \citep[e.g.,][]{Pessa2021A&A...650A.134P...PHANGS}. In particular, at low redshifts there exist relationships between the surface densities of gas mass ($\Sigma_{\rm gas}$), stellar mass ($\Sigma_*$), and star formation rate ($\Sigma_{\rm SFR}$) on sub-kiloparsec scales \citep[see][for a review]{Sanchez20}. These relationships are inherently linked to the evolution of galaxies, as they describe the formation of stellar populations. The relationship between $\Sigma_*$ and $\Sigma_{\rm SFR}$ is often called the resolved star formation main sequence \citep[rSFMS; e.g.,][]{Cano-Diaz16, Maragkoudakis16,Abdurro'uf17,Liu_2018,Trayford19,Erroz-Ferrer, Medling2018MNRAS.475.5194M, Lin2019ApJ...884L..33L,Bluck2020a,Pessa2021A&A...650A.134P...PHANGS,Barrera-Ballesteros2021ApJ...909..131B} and is the main focus of this Letter. However, we also present the resolved gas main sequence (rGMS, i.e., $\Sigma_*-\Sigma_{\rm gas}$ relation) and resolved Kennicutt-Schmidt law (rKS, i.e., $\Sigma_{\rm gas}-\Sigma_{\rm SFR}$ relation) for simulated galaxies.

Scaling relationships play an important role in cosmological hydrodynamic simulations of galaxy evolution, since unconstrained parameters can be calibrated to reproduce select relationships. For example, the IllustrisTNG simulations \citep{Nelson2018,2018MNRAS.475..648P,2018MNRAS.475..676S,2018MNRAS.477.1206N,2018MNRAS.480.5113M,Nelson2019a} were calibrated to reproduce the stellar-to-halo mass, stellar mass -- stellar size, and black hole -- galaxy mass relations. Relationships that are not used to calibrate a simulation can, instead, be used to validate it. For example, in \cite{McDonough2023} we showed that the $100-$Mpc box of the IllustrisTNG suite (TNG100) broadly reproduced the rSFMS with a slope in agreement with that found by \cite[][]{Bluck2020a} for MaNGA \citep{MaNGA} galaxies.

In general, scaling relations have broad utility in extragalactic astrophysics. For observational measurements that are challenging to make, scaling relations can be used to identify appropriate proxies, e.g., inferring black hole mass from central velocity dispersion \citep[e.g.,][]{Gebhardt2000ApJ...539L..13G}. Semi-analytic models often incorporate scaling relations, for example: populating simulated dark matter halos with galaxies given the halo-to-stellar mass relation \citep[see][for a review]{DeLucia2019Galax...7...56D}. Further, scaling relations arise due to physical processes, and therefore their shapes encode physically significant information. For example, \cite{Schmidt1959ApJ...129..243S} demonstrated a power-law relationship between SFR and the surface density of gas. \cite{Kennicutt1989ApJ...344..685K} and \cite{Kennicutt1998ApJ...498..541K} expanded on this work, finding that the SFR was also regulated by the total local density which sets the conditions for gravitational collapse. Given that the timescale for gravitational instabilities to grow in a parcel of gas will scale with local density, a star formation law should follow the form: $\rho_{\rm SFR} \propto \rho_{\rm gas}^{1.5}$ above some threshold density \citep{Kennicutt1998ApJ...498..541K}. More recently, \cite{Shi2011ApJ...733...87S} proposed an extension of the rKS law with $\Sigma_{\rm SFR}$ being dependent on both $\Sigma_{\rm gas}$ and $\Sigma_*$. Many works have found that the rKS holds only for molecular gas \citep[e.g.,][]{Kennicutt2007ApJ...671..333K}. A full review of theoretical predictions for the rKS and similar relations is beyond the scope of this Letter, but in general the slopes of these power-law relations are predicted by theory and can be constrained with observational measurements. 



As both simulations and observations improve in resolution and complexity, a need for accurate and unbiased quantitative comparisons of scaling relations arises if we are to constrain theory further. Of particular interest is whether the physical properties that drive spatially resolved scaling relations are redshift-dependent, as has been found for some global relations, e.g., the global SFMS \citep{Speagle_2014}. Unfortunately, differences in galaxy sample selection, spatial resolution, and techniques for measuring scaling relationships have made such comparisons challenging. Measurements of the rSFMS and rKS can be further affected by the criteria for pre-selection of star-forming regions and the indicator used to measure SFR. The slopes of the rSFMS presented in the literature range from $\sim 0.5$ to $\sim 1$, with reported confidence intervals ranging from $\sim 0.0002 - 0.2$ \citep[e.g.,][]{Cano-Diaz16,Maragkoudakis16,Abdurro'uf17,Liu_2018,Trayford19,Erroz-Ferrer,Bluck2020a,Hani20,Pessa2021A&A...650A.134P...PHANGS,McDonough2023}. Using spatially resolved maps of simulated galaxies from the FIRE project \citep{FIRE-2}, \cite{Hani20} found that the rSFMS slope depends on the fitting method, star formation timescales, and spatial resolution. 

The rSFMS may be a byproduct of two more fundamental relations \citep[e.g.,][]{Baker2023MNRAS.518.4767B}: the molecular gas main sequence (i.e., $\Sigma_{H_2} -\Sigma_*$) and the molecular rKS law (i.e., $\Sigma_{H_2}-\Sigma_{\rm SFR}$). However, the rSFMS remains useful due to the challenges associated with obtaining molecular gas measurements for large observational samples (i.e., the need for millimeter wavelengths). Even in simulation space, the rSFMS requires only the analysis of stellar particles, while investigations of the other relations requires analysis of both stellar and gas particles. Thus, a consistent measurement of the rSFMS slope and other scaling relations is desirable for validation of simulation results and for probing the physics that drive star formation. 

Many studies in the literature use some variation of least squares to fit a power law to their samples of spectral pixels (i.e., spaxels) or simulated analogs, usually done in logarithmic space to simplify the problem. Typically, some criteria are used to pre-select star-forming spaxels and a line is then fit to either the full sample or the median of the sample in bins of stellar mass. Mathematically, a least squares fit to a full sample returns a line that traces the conditional mean (e.g., E$[\Sigma_{\rm SFR}|\Sigma_*]$). This method is not ideal because spaxels are not normally distributed about the rSFMS; as regions of galaxies quench, the SFR of spaxels will decrease while stellar mass remains roughly fixed. Fitting a line to the median of the data (i.e., identifying a linear approximation to the conditional median) limits the influence of quenched regions, although it is still biased by non-normal distributions. The introduction of pre-selection criteria for star-forming spaxels may limit the data to the regime where it is normally distributed, but is hard to apply consistently across studies, especially those using different SFR tracers. More broadly, any scaling relations fit with least squares can be biased by the presence of outliers, non-normal scatter and other limitations on the range of data covered (e.g., magnitude- or mass- limited samples).

\cite{Renzini15} describe an objective definition for the global star-forming main sequence (SFMS; i.e., $M_*-{\rm SFR}$ relationship) based on the ``ridge line'' of data in 3D $M_*-{\rm SFR}$ plots. The ridge line method works by fitting a line to the mode of $\Sigma_{\rm SFR}$ in bins of $\Sigma_*$ (i.e., the conditional mode). This method is intended to ameliorate biases introduced by the presence of quenching spaxels without pre-selecting star-forming systems. Mathematically, the ridge line accomplishes this because it returns a function that traces the maximum of the conditional probability distribution function (e.g., max[P$(\Sigma_{\rm SFR}|\Sigma_*)$]). The ridge line method has been adopted to measure the rSFMS in both observations \citep[e.g.,][]{Abdurro'uf17} and simulations \citep{McDonough2023}. 

In this Letter, we investigate how different methodologies for obtaining maps of $\Sigma_*$ and $\Sigma_{\rm SFR}$ from galaxies in the TNG100 simulation affect the resulting rSFMS. We demonstrate that these effects are minimized when the slope is obtained by fitting a double linear function to the ridge line. Our ridge line method builds on the approach proposed by \cite{Renzini15}, and we hope to further facilitate the adoption of this technique by releasing the open-source Python package \texttt{ScaleRPy} \citep{mcdonough_2025_scalerpy}. In Section \ref{sec:data}, we discuss the TNG simulations and our methodology for obtaining spatially-resolved data. We also present the rSFMS, rGMS, and rKS for TNG100 galaxies. In Section \ref{sec:results:fit}, we present our results for rSFMS, rKS, and rGMS fits obtained with different fitting techniques. In Section \ref{sec:results:res} and \ref{sec:results:smooth}, we compare ridge line fits of the rSFMS from galaxy maps generated at different spatial resolutions and with different adaptive smoothing scales, respectively. In Section \ref{sec:summary}, we summarize our results and present our prescription for measuring unbiased, objective slopes of scaling relations. To ensure the reproducibility of this work, the data and Jupyter notebook containing the analysis software is preserved on Zenodo: \dataset[doi: 10.5281/zenodo.15047580]{\doi{10.5281/zenodo.15047580}}. The code repository for \texttt{ScaleRPy} is hosted on Github\footnote{\texttt{ScaleRPy} codebase: \url{https://github.com/bryannemcd/ScaleRPy}.} and the specific version of \texttt{ScaleRPy} used for this Letter is permanently archived on Zenodo \citep{mcdonough_2025_scalerpy}.

\section{Data and Methods} \label{sec:data}

Our galaxy sample is identical to that of \cite{McDonough2023} and \cite{McDonough2025}, and was drawn from the TNG100 simulation. TNG100 used the IllustrisTNG model of galaxy formation, which adopted a \cite{Planck2016} cosmology. TNG100 has a comoving box length of $75 h^{-1}\textrm{Mpc} $ and a target baryonic particle mass of $\sim 10^6 h^{-1}M_\odot $. The IllustrisTNG model was calibrated such that at $z=0$ it reproduces the observed stellar mass function, the total gas mass of massive groups, and the following scaling relations: stellar-to-halo mass, stellar mass to stellar size, and black hole to galaxy mass \citep{2018MNRAS.475..648P}. Additionally, the model was calibrated to reproduce the shape of the cosmic star formation rate density as a function of redshift out to $z=10$. The TNG simulations use a star formation prescription \citep{Springel2003MNRAS.339..289S} with free parameters tuned to reproduce the theoretical rKS slope of $1.5$ proposed by \cite{Kennicutt1998ApJ...498..541K}.

Our galaxy sample selection criteria are described in full in \cite{McDonough2023} and here we provide only a brief summary. Specifically, we ensure sufficient spatial resolution by limiting our sample to galaxies with $R_e > 4$ kpc and more than $1000$ stellar particles within $2R_e$, where $R_e$ is the $r$-band half-light radius. We obtain galaxy magnitudes and radii from the SDSS Photometry, Colors, and Mock Fiber Spectra \citep{Nelson2018} and Stellar Projected Sizes \citep{Genel2018} catalogs, respectively. These criteria result in a sample of $6197$ TNG100 galaxies, including centrals and satellites. We compute time-averaged SFRs using stellar particles formed in the last $20$ and $100$ Myrs. Unlike \cite{McDonough2023}, here we do not include spaxels for which no stellar particles were formed in the last $100$ Myrs. These are comparable to observational spaxels where a measurement of SFR cannot be made due to poor signal-to-noise. 

As in \cite{McDonough2023}, we obtain spatially-resolved maps of stellar mass and SFR using a cubic spline kernel to project stellar particles onto a 2D plane that corresponds to a face-on projection of the galaxy. The 2D cubic spline kernel is: 
\begin{equation}
    W(r, h_{sml}) = \frac{40}{7\pi} \frac{1}{h_{sml}^2}
    \begin{cases}
        1 - 6q^2 + 6q^3     & \text{if } 0 \leq q \leq \frac{1}{2}, \\
        2(1-q)^3            & \text{if } \frac{1}{2} < q \leq 1, \\
        0                   & \text{if } q >1,
    \end{cases}
    \label{eq:SPH}
\end{equation}
where $r$ is the radius a particle is smoothed over, $h_{sml}$ is the distance between a particle and its Nth nearest neighbor of the same particle type, and $q \equiv r/h_{sml}$. For this work, we have also generated maps of total gas mass using the same kernel. 

We choose as our fiducial parameters those used in \cite{McDonough2023}, which generated 2D maps with a spatial resolution of $500 h^{-1} {\rm pc}$ ($\sim 750$ pc) and a value of $N=64$ for the cubic spline kernel. The fiducial spatial resolution was selected to be approximately comparable to the resolution of MaNGA data used by \cite{Bluck2020a} and \citep{Bluck2020b}. The fiducial value of $N$ was selected to improve the relative coverage of star-forming regions. Additional maps of $\Sigma_*$ and $\Sigma_{\rm SFR}$ were generated at different spatial resolutions and different values of $N$ to explore the effect of these parameters on the resulting fit. Using $N=64$ we construct maps with spatial resolutions of $500$ pc, $\sim 750$ pc, $1$ kpc, and $\sim 1.5$ kpc. At a spatial resolution of $\sim 750$ pc, we also construct maps using $N = 8$, $16$, $24$, $32$, and $64$. The values of $N$ that we explore here were inspired by the range of $N$ over which \cite{Rodriguez-Gomez2019_morph} found their morphological measurements of TNG100 galaxies to be insensitive. We note that this parameter does not have a direct analog in observed data, but the choice of $N$ for simulated galaxies impacts the spatial extent of star-forming regions and the patchiness of stellar mass maps. Pixels in 2D maps of galaxies from data obtained with integral field spectroscopy (IFS) are often referred to as spectral pixels or `spaxels.' While the pixels in our 2D maps are not true spaxels, we refer to them as such for the sake of clarity. We limit our analysis to spaxels with $\Sigma_{\rm SFR}>10^6 M_\odot {\rm kpc}^2$, consistent with \cite{Bluck2020b} and \cite{McDonough2023}. For the rKS and rGMS relations, we further limit our analysis to spaxels with $\Sigma_{\rm gas}>10^6 M_\odot {\rm kpc}^2$, the approximate observational limit of recent high-resolution surveys of nearby galaxies \citep[e.g.,][]{Pessa2021A&A...650A.134P...PHANGS}.

In Section \ref{sec:results:fit}, we present the rSFMS, rKS, and rGMS slopes that we obtain using five different methods to fit spatially resolved data: [1] ridge line: our fiducial approach described below; [2] OLS: ordinary least squares applied to the spaxel data in $\log_{10}$ space; [3] LS-1PL: non-linear least squares to fit the spaxel data to a single power law; [4] LS-2PL: non-linear least squares to fit the spaxel data to a double power law; and [5] LS-2PL-log: non-linear least squares to fit the spaxel data to a double linear in $\log_{10}$ space. While many studies have fit the spatially resolved scaling relations with a single power law,  several studies have found that there is a point at which the global SFMS turns over \citep[e.g.,][]{Renzini15,Popesso2019MNRAS.483.3213P,Mancini2019MNRAS.489.1265M}, and the bend also appears in the rSFMS \citep[e.g.,][]{Abdurro'uf17,McDonough2023}. A bend in the rKS is also expected at the density threshold for star formation to occur \citep{Kennicutt1989ApJ...344..685K}. Any rSFMS or rKS fits made to a single power law function are thus either biased by these transitions or must pre-select the regime where the relationship is purely linear in logarithmic space. For this reason, methods [1], [4], and [5] fit to the functional form of a double power law, and we compare this to a single power law with methods [2] and [3]. We further demonstrate the difference between fitting in linear space$-$ methods [2] and [4]$-$and logarithmic space$-$ methods[1], [3], and [5]. As we will show, this difference can be significant.


We use the open-source \texttt{ScaleRPy} Python package to identify and fit the ridge line described by \cite{Renzini15}. In each bin of $\log_{10} \Sigma_*$, we use a Gaussian kernel density estimator \citep[KDE; e.g.,][]{Scott1992-sv} to identify the mode of $\log_{10}\Sigma_{\rm SFR}$. The ridge of our data is defined as the collection of points that correspond to the centers of the $\log_{10}\Sigma_*$ bins and the mode of the KDE of $\log_{10}\Sigma_{\rm SFR}$ in those bins. Unlike a mean or median, the mode of a data set in linear space will be identical to the mode of a data set in $\log_{10}$ space. We choose to perform the ridge line fit in $\log_{10}$ space because the KDE software we use assumes a fixed bandwidth, which works poorly for data that span many orders of magnitude. We have tested that fitting the identified ridge points with least squares in linear space does not significantly affect our results. Uncertainties on the ridge points are estimated as the KDE bandwidth for a given bin of $\Sigma_*$ and determined via Scott's rule \citep{Scottsrule}, which considers the number of data points and their standard deviation from a normal distribution. The \textnormal{rSFMS and rGMS} ridge line and fits are computed over $10^6 < \Sigma_*/[M_\odot {\rm kpc}^2] < 10^{9}$, where there is sufficient data to obtain a KDE for each mass bin in the rSFMS. The ridge line is fit with a double power law parameterized as: 
\begin{equation}
    \centering
    \log_{10} \Sigma_{\rm SFR,MS} = 
    \begin{cases}
        \gamma \log_{10} \Sigma_* + y_0     & \text{if } \Sigma_* < \Sigma_*^{\rm to}, \\
        \gamma' \log_{10} \Sigma_* + y_0'           & \text{if } \Sigma_* >  \Sigma_*^{\rm to}, 
    \end{cases}
    \label{ch3:eq:doublefit}
\end{equation}
where $\gamma$ is the rSFMS slope, $y_0$ is the y-axis zero point, $\gamma'$ is the slope of the high-mass end of the rSFMS, $\Sigma_*^{\rm to}$ is the value of $\Sigma_*$ where the main sequence turns over, and $y_0' \equiv (\gamma - \gamma') \log_{10}\Sigma_*^{\rm to} +y_0$. A similar parameterization is used to fit the rKS and the rGMS. 

\begin{figure}[ht!]
	\centering
    \subfloat[rSFMS\label{fig:rSFMS}]{%
        \includegraphics[width=0.49\linewidth]{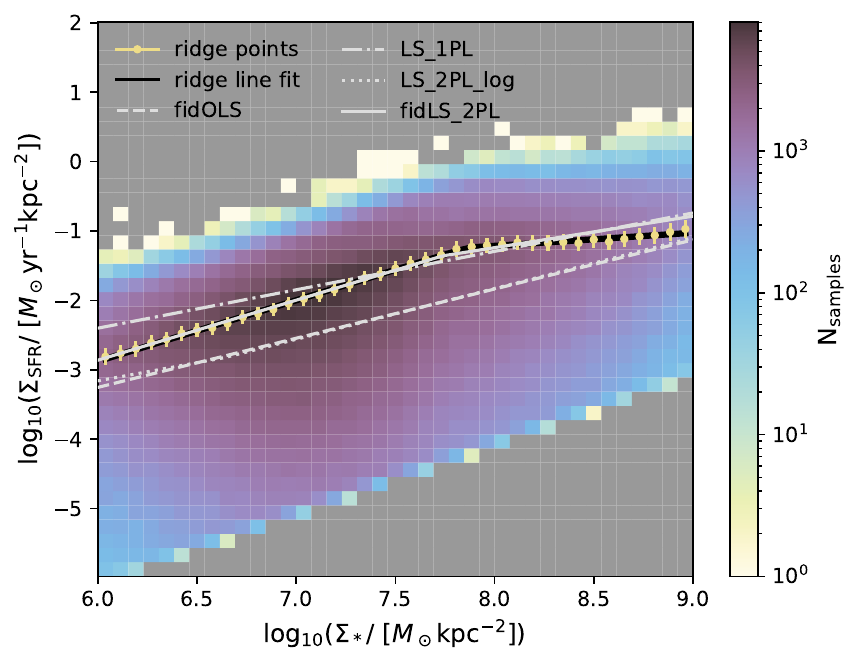}%
    }%
    ~ 
    \subfloat[rKS\label{fig:rKS}]{%
        \includegraphics[width=0.49\linewidth]{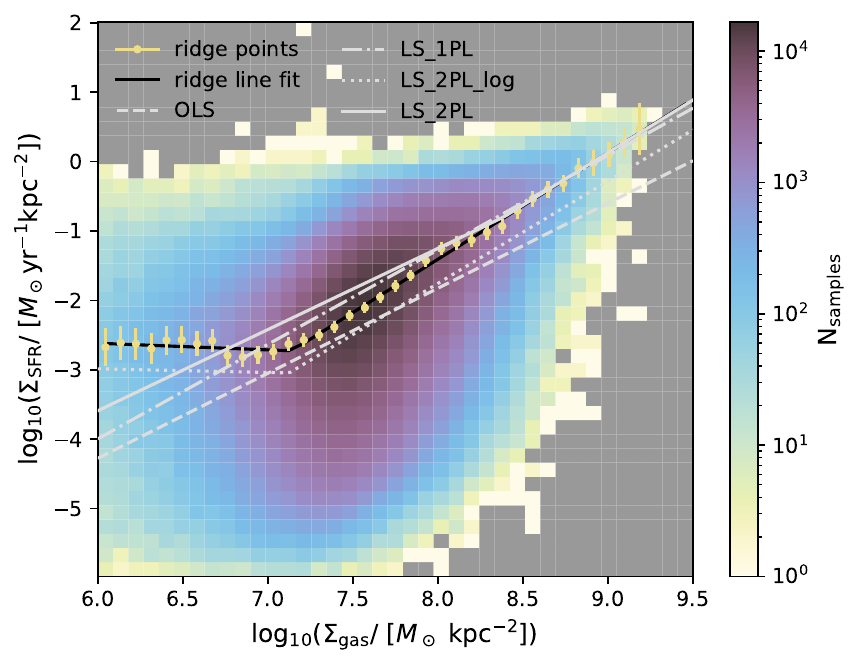}%
    }\\
    \medskip 
    \subfloat[rGMS\label{fig:rGMS}]{%
        \includegraphics[width=0.475\linewidth]{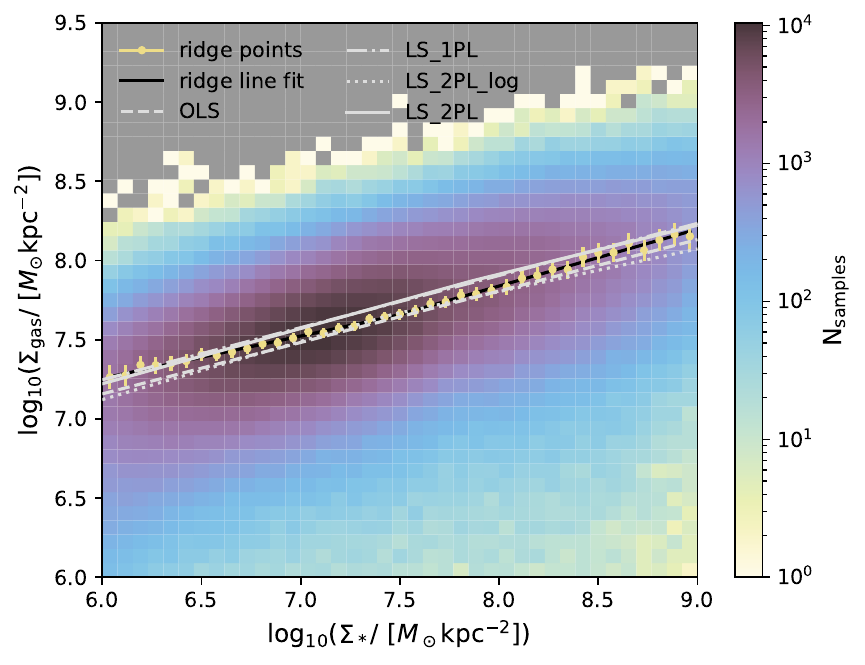}%
    }%
    \subfloat[\label{fig:3D}]{%
            \includegraphics[scale=0.5]{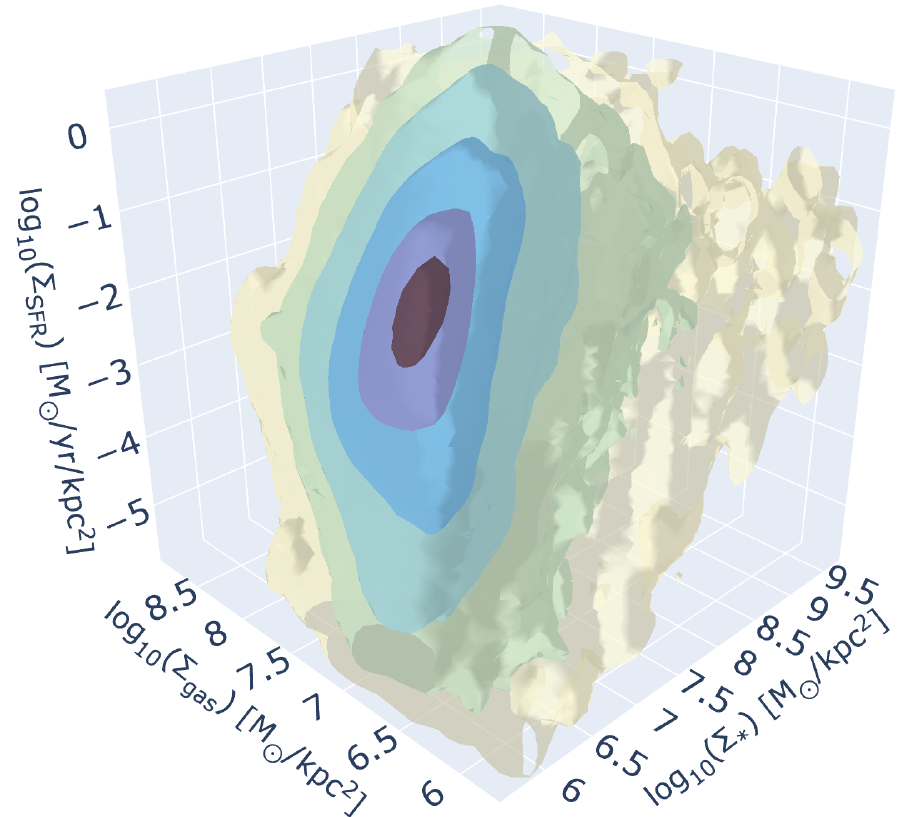}
    }
    
    \caption{Panels a-c: The rSFMS (a), rKS (b), and rGMS (c) scaling relations from our fiducial maps, which demonstrate the ridge line and ordinary least squares fitting techniques. The 2D histogram shows the number of spaxels (i.e., samples) in bins of values on the x and y axes. Yellow points indicate the `ridge' or \textnormal{conditional} mode of the data, and the black line shows the double linear fit to those ridges. Uncertainties on the ridge points are estimated as the bandwidth of the KDE obtained with Scott's rule \citep{Scottsrule}. The gray lines show least squares fits to a single power law (dashed and dot-dashed lines) and to a double power law (dotted and solid lines) using the full data set in linear space (dot-dashed and solid) and in $\log_{10}$ space (dashed and dotted). Panel d: A $3$D visualization of the relationship between $\Sigma_*$, $\Sigma_{\rm gas}$, and $\Sigma_{\rm SFR}$. Iso-density contours are drawn at the $50^{\rm th}$, $70^{\rm th}$, $85^{\rm th}$, $95^{\rm th}$, $99.9^{\rm th}$ percentiles. In the html version of the published Letter, this panel is available as an interactive figure, where the $3$D visualization can be rotated.  \label{fig:fid_ex}}
\end{figure}

In Figure \ref{fig:fid_ex}, we compare the fits obtained with least squares (gray) and the ridge line (black) methods for the distributions of $\Sigma_*-\Sigma_{\rm SFR}$ (Figure \ref{fig:rSFMS}), $\Sigma_{\rm gas}-\Sigma_{\rm SFR}$ (Figure \ref{fig:rKS}), and $\Sigma_*-\Sigma_{\rm gas}$ (Figure \ref{fig:rGMS}) from our fiducial maps. The yellow points indicate the ridges in each bin of $\Sigma_*$. The dashed lines are the fits obtained with OLS, dot-dashed lines are fits obtained with LS-1PL, dotted lines are fits obtained with LS-2PL-log, and the solid gray lines are fits obtained with LS-2PL. In Figure \ref{fig:3D}, we visualize the distribution of spaxel data in 3D space with iso-density contours, to highlight the relationship between all three variables. 
Throughout, we use the $R^2$ statistic, also known as the coefficient of determination, to measure the goodness of our fits. 
A value of $R^2 =1$ means that the model fits the data perfectly. 
Comparison of the $R^2$ statistic for ridge line and least squares fits are not necessarily straightforward, since the ridge line technique fits to far fewer points. Thus, the variation of data from least squares fits is naturally greater. However, this is one of the advantages of the ridge line technique: it first identifies what it is that should be fit (the maximum of the conditional probability density function) before fitting a relation.

\section{Results} \label{sec:results}

\begin{deluxetable}{lccccccc}
    \tabletypesize{\footnotesize}
    \tablecolumns{7}
    \tablewidth{0pt}
    \tablecaption{Parameters of fits to the $z=0$ resolved star formation main sequence recovered from maps of TNG100 galaxies generated with different methods.
    \label{tab:rSFMS_runs}}
    \tablehead{\colhead{Name} & \colhead{Resolution (pc)} & \colhead{N} & \colhead{$\gamma$} & \colhead{$y_0$} & \colhead{$\gamma'$} & \colhead{$\log_{10}[\Sigma_*^{\rm to} M_\odot^{-1}{\rm kpc}^2]$} & \colhead{${\rm R}^2$} }
    \startdata
    fidOLS & $\sim 0.75$ kpc & $64$ & $0.709 \pm 0.001$ & $-7.509 \pm 0.008$ & N/A & N/A & $0.2170$\\
    fidLS-1PL & $\sim 0.75$ kpc & $64$ & $0.554 \pm 0.001$ & $-5.722 \pm 0.009$ & N/A & N/A & $0.1776$\\
    fidLS-2PL & $\sim 0.75$ kpc & $64$ & $0.865 \pm 0.006$ & $-8.06 \pm 0.05$ &  $0.454 \pm 0.002$ & $7.738 \pm 0.007$ & $0.1813$\\
    fidLS2-log & $\sim 0.75$ kpc & $64$ & $0.52 \pm 0.01$ & $-6.31 \pm 0.06 $ &  $0.7323 \pm 0.002$ & $6.64 \pm 0.02$ & $0.217$\\
    fiducial  & $\sim 0.75$ kpc & $64 $ & $0.88 \pm 0.01$ & $-8.15 \pm 0.07$  & $0.19 \pm 0.02$ & $7.86 \pm 0.02$ & $0.9984$\\
    N32 & $\sim 0.75$ kpc & $32 $ & $0.90 \pm 0.01$ & $-8.07 \pm 0.08$ & $0.16 \pm 0.01$ & $7.58 \pm 0.02$ & $0.9979$ \\
    N24 & $\sim 0.75$ kpc & $24 $ & $0.91 \pm 0.01$ & $-8.07 \pm 0.08$ & $0.16 \pm 0.01$ & $7.58 \pm 0.02$ & $0.9979$ \\
    N16 & $\sim 0.75$ kpc & $16 $ & $0.91 \pm 0.02$ & $-8.0 \pm 0.1$ & $0.15 \pm 0.01$ & $7.32 \pm 0.02$ & $0.9974$ \\
    N8 & $\sim 0.75$ kpc & $8 $ & $0.86 \pm 0.02$ & $-7.4 \pm 0.2$ & $0.13 \pm 0.01$ & $7.12 \pm 0.02$ & $0.9952$ \\
    res500 & $0.5$ kpc & $64 $ & $0.86 \pm 0.01$ & $-8.02 \pm 0.06$ & $0.12 \pm 0.05$ & $8.25 \pm 0.03$ & $0.9982$\\
    res1000 & $1$ kpc & $64 $ & $0.86 \pm 0.03$ & $-8.0 \pm 0.2$ & $0.31 \pm 0.03$ & $7.50 \pm 0.07$ & $0.9866$\\
    res1500 & $\sim 1.5$ kpc & $64 $ & $0.86 \pm 0.1$ & $-8.1 \pm 0.7$ & $0.42 \pm 0.04$ & $7.0 \pm 0.2$ & $0.9552$
    \enddata
    
    \tablecomments{Fits are done with the ridge line method except for the fidOLS and fidLS cases.}  

\end{deluxetable}

\subsection{Fitting Method} \label{sec:results:fit}


In this section, we compare results from our ridge line fits to variations of least squares fitting techniques. In Table~\ref{tab:rSFMS_runs} we report all parameters for the rSFMS fits. We summarize the interesting results for the rSFMS and rKS below, but for precise details of each fit we refer readers to the computational notebook archived on Zenodo\footnote{\dataset[doi: 10.5281/zenodo.15047580]{\doi{10.5281/zenodo.15047580}}; The notebook also includes least squares fits of the rSFMS at other resolutions.}  

In Figure \ref{fig:fid_ex}, we compare fits to the rSFMS with the ridge line method to fits obtained with least squares. When the rSFMS is fit with a single power law, the slopes ($0.709 \pm 0.001$ for fidOLS and $0.554\pm0.001$ for LS-1PL) obtained are significantly shallower than that obtained with the ridge line method ($0.88\pm0.01$) \textnormal{at $\log_{10}\Sigma_*<7.86\pm 0.02$} . Where least squares are used to fit the rSFMS in $\log_{10}$ space (fidOLS, LS-2PL-log), they are biased toward lower values of $\Sigma_{\rm SFR}$. The fidLS-2PL fit comes closest to identifying the ridge line, with a low-mass slope and intercept in agreement with the ridge line fit. However, the turnover density from fidLS-2PL ($7.738 \pm 0.007$) is lower than that from the ridge line ($7.87 \pm 0.02$) and the slopes at high-densities also disagree. Although it more closely follows the ridge points (i.e., conditional modes) compared to other least squares fits, the fidLS-2PL method does not result in as good a fit to the ridge points as the ridge line method ($R^2 =0.6828$ compared to $R^2=0.9984$). 

The significant differences between the least squares fits and the ridge line fit are unsurprising. For relationships with $\Sigma_{\rm SFR}$, the main advantage of the ridge line method is that it is not biased by the presence of spaxels naturally transitioning off the main sequence. The $R^2$ statistic shows that the least squares models do a poor job of describing the variation of $\Sigma_{\rm SFR}$ with $\Sigma_*$. For the \textnormal{least squares fits, $R^2\sim 0.2$}. However, with $\rm R^2 = 0.9984 \approx 1$, the ridge line method results in a model that fits the data well. Although it may be possible to improve $\rm R^2$ for the least squares cases by imposing additional criteria \textnormal{such as priors and limiting} the sample to spaxels that are actively star-forming, this would defeat the goal of obtaining an objective and broadly comparable rSFMS slope. 

\textnormal{We now turn to analysis of fits to the rKS and rGMS relations from our fiducial maps. We note that $\Sigma_{\rm gas}$ is the total surface density of all gas, rather than a molecular subset (e.g., $\Sigma_{H_2}$).}

\textnormal{In Figure \ref{fig:rKS}, we compare fits to the rKS relation obtained with a double linear ridge line fit and least squares fits. At $\Sigma_{\rm gas}\lesssim10^7 M_\odot {\rm kpc}^{-2}$, there appears to be essentially no significant relationship between $\Sigma_{\rm gas}$ and $\Sigma_{\rm SFR}$ (the slope of the ridge line fit at $\Sigma_{\rm gas}<10^{7.14\pm0.03} M_\odot {\rm kpc}^{-2}$ is $-0.10\pm0.08$ and data points are approximately normally distributed over a large range). This is generally consistent with theoretical predictions of a density threshold around $\Sigma_{\rm gas} \sim 10^7M_\odot {\rm kpc}^{-2}$ that must be met for star formation to occur \citep[e.g.,][]{Kennicutt1989ApJ...344..685K,Springel2003MNRAS.339..289S}. At higher gas surface densities, the ridge line is fit by a line with a slope of $1.54\pm0.02$. The slope obtained with least squares range from $1.226\pm0.002$ (for OLS) to $1.516 \pm0.00
4$ (for LS-2PL at high $\Sigma_{\rm gas}$). Our analysis of this relationship in TNG100 implies that fits of the rKS to a single power law  will be affected by the availability of data for gas at low surface densities. The $R^2$ statistic for the ridge line fit to the rKS is $0.9967$, compared to $0.2129$ for the OLS fit.}

\textnormal{Unlike the other relations we identify here, the prescriptions to estimate subgrid star formation in the TNG simulations were specifically tuned to reproduce the rKS relation with a slope of $1.5$ at $\Sigma_{\rm gas}\gtrsim 10^7\; M_\odot \;{\rm kpc}^2$ \citep{Springel2003MNRAS.339..289S}. This tuning was not trivial, and the relationship that results from solving the equations which govern star formation in the \cite{Springel2003MNRAS.339..289S} model has a slope that is not reported but appears slightly steeper than $1.5$. Because this model is used to calculate the SFR for star-forming gas particles in the TNG100 simulation, any rKS relationship recovered from that simulation should have a slope in agreement with that model. The rKS slope we identify with the ridge line method ($1.54 \pm 0.02$), as well as those obtained with least squares fits to double power laws ($1.476\pm0.002$ and $1.516\pm0.004$ for LS-2PL and LS-2PL-log, respectively) are generally consistent with a slope of $1.5\pm 0.1$, but we leave a more rigorous statistical comparison for a later work. According to the \cite{Springel2003MNRAS.339..289S}, the rKS should turn over at $\Sigma_{\rm gas} \sim 10^7 M_\odot {\rm kpc}^2 M_\odot {\rm kpc}^{-2}$. The ridge line method returns a turnover density of $10^{7.14\pm 0.03}$ while LS-2PL-log and LS-2PL return turnover densities of $10^{7.126\pm 0.002} M_\odot {\rm kpc}^{-2}$ and $10^{8.389\pm 0.008} M_\odot {\rm kpc}^{-2}$, respectively. Thus, only the ridge line method and a least squares fit to a double power law in $\log_{10}$ space returns parameters in agreement with the star formation models built into the TNG100 simulation. The only significant difference between the ridge line fit and LS-2PL-log for the rKS is in the normalization of the relation. }

\textnormal{In Figure \ref{fig:rGMS}, we show the distribution of $\Sigma_*$ and $\Sigma_{\rm gas}$.  Within the bounds of $\Sigma_*$ and $\Sigma_{\rm gas}$ that we explore here, fitting the ridge line with a double power law has little effect on the resulting rGMS slope compared to fitting to a single power law, with a difference of $\sim0.1$ between the slopes at low- and high- density for all double power law fits. The slopes we find from all methods range from $0.27$ to $0.37$. Normalization is the main difference in the rGMS fits done in linear and logarithmic space, as expected. The $R^2$ values for the least squares fits to the rGMS are better than those for the least squares fits to the rKS or rSFMS but still objectively poor, with $R^2\approx0.32$ compared to $R^2 = 0.997$ for the ridge line method.
 }

In the following subsections, we fit the rSFMS with a double \textnormal{power law }using the ridge line method.



\subsection{Spaxel Resolution} \label{sec:results:res}
\begin{figure}[htp!]
	\centering
    \includegraphics[width=\linewidth]{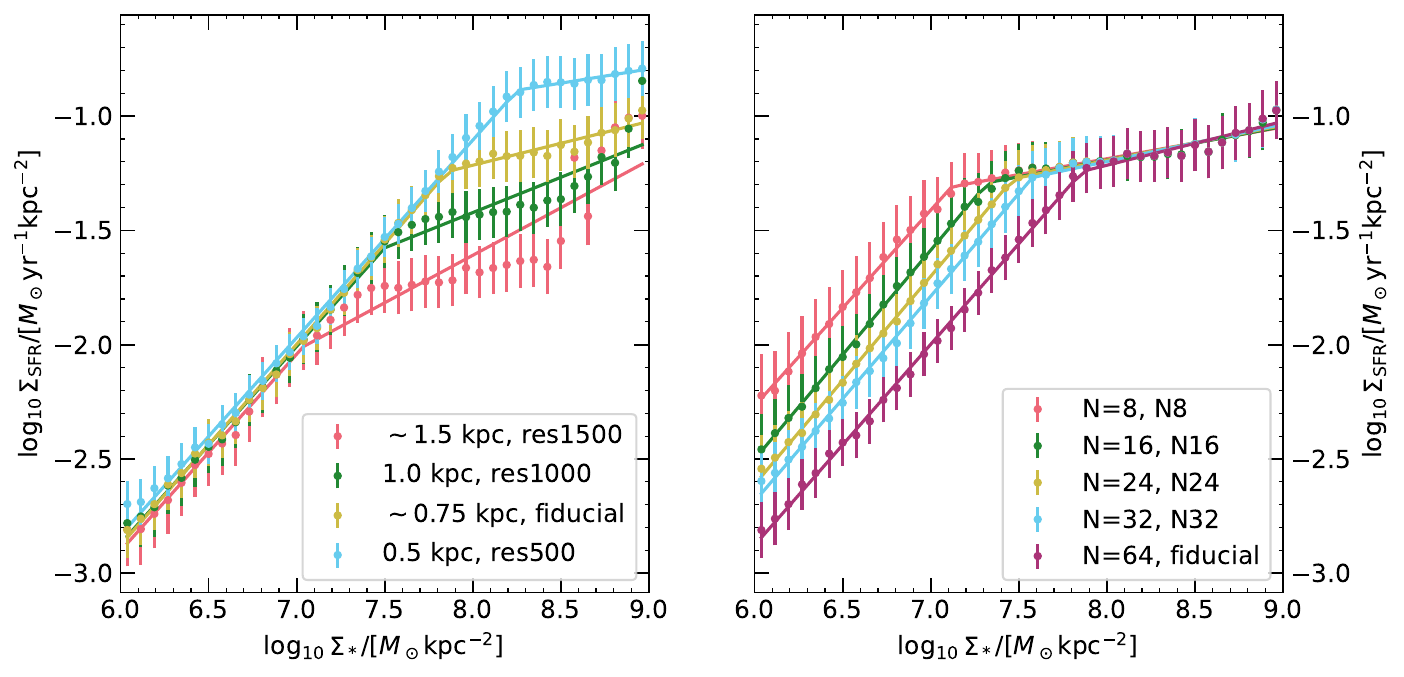}

    \caption{Comparison of the modes of spaxels in the star-forming portion of the $\Sigma_* - \Sigma_{\rm SFR}$ diagram derived from maps with different spatial resolutions (left) and different smoothing scales (right). 
    Points represent the ridges identified in each bin of $\Sigma_*$ and lines of the same color represent the fit to those ridges. The uncertainties are approximated as the KDE bandwidth estimated via Scott's rule. \label{fig:mode_comp}}
    
\end{figure}

In Figure \ref{fig:mode_comp}, we present the ridge lines derived from maps with different spatial resolutions (left) and different numbers of nearest neighbors for the smoothing radius (right). 

Previous results \citep{Hani20} have indicated that the spatial resolution of the surface density maps may affect the slope obtained for the rSFMS. In Figure \ref{fig:mode_comp} (left), we plot the ridge lines from the $2$D distribution in the $\Sigma_* - \Sigma_{\rm SFR}$ diagram of star-forming spaxels taken from maps with the following spaxel resolutions $-$ blue: $0.5 {\rm kpc}$ ($\sim 0.34 h^{-1}{\rm kpc}$), yellow: $\sim 0.74{\rm kpc}$ ($0.5 h^{-1}{\rm kpc}$), green: $1 {\rm kpc}$ ($\approx 0.68 h^{-1}{\rm kpc}$), and pink: $\sim 1.5 {\rm kpc}$ ($1 h^{-1}{\rm kpc}$). Points show the `ridge line' or modes of the distribution of $\Sigma_{\rm SFR}$ in bins of $\Sigma_*$, while the solid lines show the fits to the ridge lines. Altering the resolution of the maps affects the surface mass at which the main sequence turns over, but the slopes at the low-mass end agree within $1\sigma$ at all resolutions (see Table \ref{tab:rSFMS_runs}). 

\begin{figure}[ht!]

	\centering
    \includegraphics{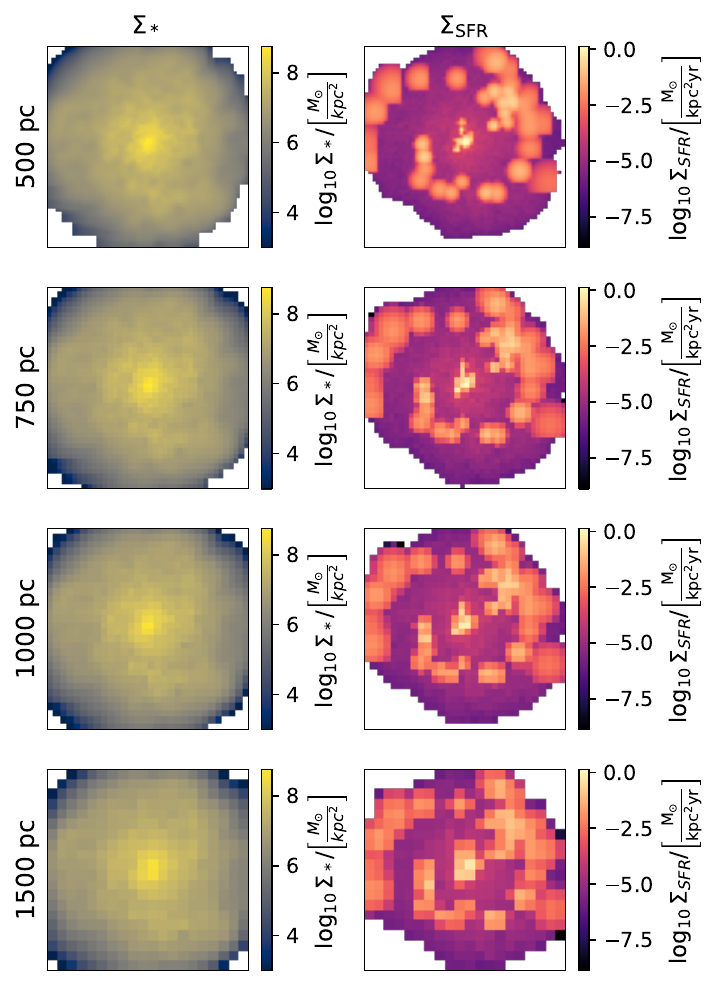}

    \caption{Maps of $\Sigma_*$ (left) and $\Sigma_{\rm SFR}$ (right) for TNG100-99-540858. Each row contains maps generated at different spaxel resolutions, as indicated on the left. 
    \label{fig:resmap_loms}}
    
\end{figure}

As the spaxel size decreases, the rSFMS turns over at higher values of $\Sigma_*$. Due to the large number of contributing stellar particles and relatively large smoothing radius ($N=64$), the stellar mass surface density should remain relatively constant with spaxel size. However, the sparser sampling of stellar particles contributing to $\Sigma_{\rm SFR}$ leads to concentrated star formation regions being spread over larger spaxels at lower resolutions, resulting in lower SFR surface densities.

Figure \ref{fig:resmap_loms} illustrates the effects of spatial resolution on maps of $\Sigma_*$ and $\Sigma_{\rm SFR}$ for a particular TNG100 galaxy, TNG100-99-540858\footnote{This identifier indicates that this galaxy is from the 99$^{\rm th}$ snapshot of TNG100, and is indexed as 540858 in the subhalo catalog.}. TNG100-99-540858 is a main-sequence, central galaxy with a stellar mass of $6.4 \times 10^{9} M_\odot$. The central star-forming region seen in Figure \ref{fig:resmap_loms} is spread over a larger area at lower resolutions. This effect contributes to the trend observed in Figure \ref{fig:mode_comp} (left).

From Table~\ref{tab:rSFMS_runs}, the value of $\rm R^2$ decreases from the fiducial value ($\rm R^2 = 0.9983$) for the res1000 and res1500 cases, which have $\rm R^2 = 0.9866$ and $0.9552$, respectively. This is likely due to the large variation in the ridge line at high masses, which we attribute to a reduction in the number of high-$\Sigma_*$ spaxels in these cases. This reduction is a result of there being fewer spaxels in low-resolution maps and that these spaxels have \textnormal{larger} surface areas. 

\subsection{Adaptive Smoothing Scale} \label{sec:results:smooth}

Equation \ref{eq:SPH} describes how simulated stellar particles are smoothed onto a grid of spaxels to determine spatially resolved \textnormal{parameters}. In this equation, particles are smoothed over a radius of $h_{sml}$ which is defined as the distance of a particle to its $N^{\rm th}$ nearest neighbor. 

In Figure \ref{fig:mode_comp} (right), we present the rSFMS ridge line that results from maps generated with $N= 8$ (pink), $16$ (green), $24$ (yellow), $32$ (blue), and $ 64$ (purple).  From Table~\ref{tab:rSFMS_runs}, the rSFMS slopes agree within $2 \sigma$ for all values of $N$ explored here. However, for $\Sigma_* < \Sigma_*^{\rm to}$, the modes are offset such that as $N$ increases, the zero point decreases. For $\Sigma_*> \Sigma_*^{\rm to}$, all cases turn over at $\Sigma_{\rm SFR} \sim 10^{-1.25} M_\odot {\rm yr}^{-1} {\rm kpc}^{-2}$, but the turnover occurs at higher $\Sigma_*$ due to the offset in normalization.

\begin{figure}[t!]
	\centering
    \includegraphics[scale=0.9]{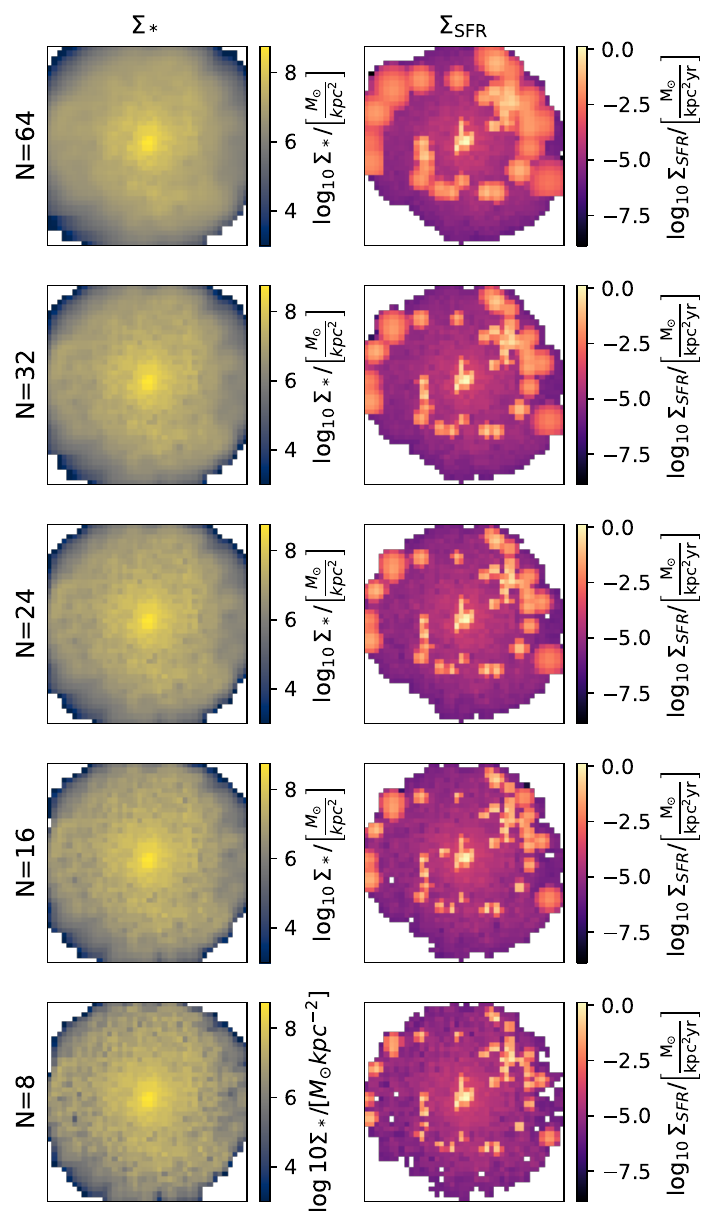}

    \caption{Maps of $\Sigma_*$ (left) and $\Sigma_{\rm SFR}$ (right) for TNG100-99-540858. Each row contains maps generated with different values of $N$,as indicated on the left.  
    \label{fig:Nmap_loms}}
    
\end{figure}

Figure \ref{fig:Nmap_loms} illustrates the effects of smoothing scale on maps of $\Sigma_*$ and $\Sigma_{\rm SFR}$ for the galaxy shown Figure \ref{fig:resmap_loms}. For small $N$ values, the maps become patchier, while the typical size of a star-forming region in the right column becomes smaller. 

As $N$ decreases, the radius over which a stellar particle is smoothed decreases. Unlike the cases in which resolution is varied, this affects both the measurements of $\Sigma_*$ (determined via the distribution of all stellar particles) and $\Sigma_{\rm SFR}$ (determined via the distribution of recently formed stellar particles). Thus, it is difficult to discern why altering $N$ has the effect seen in Figure \ref{fig:mode_comp} (right). However, the general trend of increasing $\Sigma_{\rm SFR}$ with decreasing $N$ can be explained by the concentration of stellar particles that contribute to star formation when smoothed over smaller radii. That is, higher values of $\Sigma_{\rm SFR}$ are obtained when young stellar particles are smoothed over fewer spaxels.

\section{Summary \& Discussion} \label{sec:summary}

\textnormal{In this work, we have explored different methodologies for obtaining fits to spatially resolved scaling relationships in the TNG100 simulation. We find that fits to the rSFMS and rKS can be highly sensitive to the assumed functional form of the relation and the fitting method. We propose our fiducial method, an extension of the ridge line technique presented by \cite{Renzini15}, as an objective, consistent approach to measuring scaling relations. We show that applying our ridge line method to the rSFMS for data realized at different spatial resolutions returns low-density slopes that are identical within $1\sigma$. Such insensitivity to spatial resolution is an important criteria for a fitting technqiue that is comparable across many data sets.}

\textnormal{An exhaustive exploration of possible functional forms for the rSFMS, rKS, and rGMS is outside the scope of this work. However, we have shown that the rKS and rSFMS are better described by a double, rather than single, power law. Physically, this means that star formation proceeds differently in different density regimes. In the case of the rKS, a transition occurs above some critical density for gravitational stability ($\Sigma_{\rm gas} \gtrsim 10^7 M_\odot {\rm kpc}^2$). The physical origin of a bend in the rSFMS is less clear. In Figure \ref{fig:mode_comp} we show that the turnover density is dependent on both spatial resolution and smoothing scale. At high-densities, the relationship may not be well described by a power law.} However, we demonstrate that there is a remarkable consistency in the slope at the low-density end ($\gamma$). By fitting a double power law to the ridge line of the rSFMS, we are able to parameterize the differences at the high-mass end, while obtaining an objective slope for the low-mass end of this distribution. \textnormal{For the rKS and rSFMS, fits to a single power law consistently underestimate the slope at high- and low- densities, respectively. Better results can be obtained when the least squares fits are limited to a regime that is well described by a single function. However, this introduces a subjective choice into the identification of a scaling relationship slope that limits comparability. Further, selecting a model which parameterizes the location of a regime change, such as the double power law we adopt, allows for more physically significant information to be obtained.   } 

\textnormal{The ridge line method offers significant benefits over straightforward least squares when it comes to identifying a functional form for scaling relations. Identifying the ridge line before fitting a relationship constrains the scope of the problem and minimizes the influence of the secondary processes which drive scatter from the relation. This is evident in the $R^2$ statistic for our fits. For the rSFMS, rKS, and rGMS from our fiducial maps $R^2>0.99$, while typical $R^2$ values for the least squares fits are around $0.2$. Between all of the least squares fits done across the three relationships explored in this work, the best $R^2$ value was found for the double linear fit to the rGMS in $\log_{10}$ space ($R^2 =0.36$).\footnote{All $R^2$ values and fitting results are available in the computational notebook linked above.} }

In our efforts to obtain comparable, objective measurements of the rSFMS slope, we have tried to eliminate subjective choices such as imposing bounds on the fit. However, we found it necessary to impose an upper limit on $\Sigma_*$, where the identified ridge points are no longer well described by a double \textnormal{power law}. This may be due in part to the relative sparseness of high surface mass density spaxels, resulting from the poor sampling of very high-mass galaxies in the limited volume of TNG100. The advantage of fitting a double \textnormal{power law} is that the slope at the low-mass end ($\gamma$) is not biased by our imposed upper limit. However, this is not the case for the slope at the high-mass end ($\gamma'$). It is not clear whether the high-mass end of the rSFMS is best described by a \textnormal{power law}, or what causes the change in slope at $\Sigma_*^{\rm to}$. Our results indicate that the location of the turnover point depends at least in part on spatial resolution and the sizes of star-forming regions. Further research will be necessary to understand and constrain the shape of the high-mass end of this distribution and the location of the turnover point. 

\textnormal{We identify a systematic offset toward lower normalizations in least squares fits when the relationships are fit in $\log_{10}$ rather than linear space. There are significant benefits to fitting power-law relationships in logarithmic space, given that it linearizes problem and balances the influence of data points that span many orders of magnitude. However, since least squares fits trace the conditional mean of a distribution, fits done in logarithmic space will trace the conditional \textit{geometric} mean of the distribution. This explains the consistent negative offset we observe when applying least squares to logarithmic data\footnote{Mathematically, this is proved by Jensen's inequality, which states that $\log(E[Y|X])\geq E[\log(Y)|X]$.}, which was also found by \cite{Pessa2021A&A...650A.134P...PHANGS}. Here, we cannot say definitively whether fitting should be performed before or after transformation into logarithmic space. In part, this will depend on whether the scatter from a scaling relationship is normally or log-normally distributed. In some studies, conditional means or medians have been identified in linear space and then fit in logarithmic space \citep[e.g.,][]{Medling2018MNRAS.475.5194M,Trayford19,Pessa2021A&A...650A.134P...PHANGS}, which should mitigate the normalization offset. Modes are invariant to logarithmic transformations and unaffected by asymmetric distributions of data, which may make them a better choice than means or medians. However, we have not directly compared fitting conditional means or medians to the ridge line method, although such a study would be of interest. } 

\textnormal{Defining scaling relations by the conditional modes is physically motivated. If the goal of obtaining scaling relations or `main sequences' is to describe the typical or `main' way that a galaxy or spatially resolved region behaves, that is described by the mode. In other words, scaling relations should trace the peak of the conditional probability distribution function (i.e., the conditional mode). Our justification for this statement is twofold: [1] scaling relationships arise from fundamental physical processes (e.g., the rGMS arises from gas and stars following the same gravitational potential) while deviations from these relations arises from secondary processes (e.g., star-forming regions moving off the rSFMS due to quenching driven by heating and winds) and [2] scaling relations are often used to estimate an unknown variable from a known variable (e.g., estimating black hole mass from its relationship with central velocity dispersion). Thus, objectively quantifying scaling relationships and their scatter via conditional modes is important for understanding the physical processes that drive them and is necessary for accurately using these relations to estimate unmeasured variables.}

\textnormal{As we have shown here, different approaches to fitting scaling relationships make direct comparisons between works challenging, if not impossible. Additionally, for the rKS and rGMS, we have found the relationships only for total gas density, rather than a molecular subset as is more common. However, the slopes we identify with the ridge line technique are in general agreement with the literature. At the low-density end of the rSFMS, we find a slope of $0.88\pm0.01$ while estimates of the rSFMS slope in the literature range from $\sim0.5-1$ \citep[e.g.,][]{Cano-Diaz16,Maragkoudakis16,Abdurro'uf17,Liu_2018,Trayford19,Erroz-Ferrer,Bluck2020a,Hani20,Pessa2021A&A...650A.134P...PHANGS}. We find a slope of $1.53$ at $\Sigma_{\rm gas}>10^{7.13\pm0.02}$ for the rKS, with estimates in the literature ranging from $\sim1.3-\sim3.1$ \cite{}. To our knowledge, there are not comparable measurements of the total gas main sequence in the literature, but we note that the slopes we identify are all much smaller than estimates for the \textit{molecular} gas main sequence \citep[e.g.,][]{Pessa2021A&A...650A.134P...PHANGS}. }

\textnormal{While quantitative comparisons of slopes are challenging, we can qualitatively compare our resutls for resolution dependence to other studies. We have shown that} at $\Sigma_*< \Sigma_*^{\rm to}$, the \textnormal{rSFMS} slopes we obtain with the ridge line method agree with each other within $2\sigma$ in almost all cases. The only exception is the res500 case, where the low-mass slope agrees with all others only at $3\sigma$.\footnote{We note that a resolution of $500$ pc is close to the spatial resolution limit of TNG100, which has a gravitational softening length of $\sim 180$ pc.} \textnormal{The resolution dependence of the rSFMS has also been explored by \cite{Hani20} and \cite{Pessa2021A&A...650A.134P...PHANGS}. \cite{Hani20} used ordinary least squares to identify the rSFMS for galaxies from the FIRE project}
and found that the slope became steeper as spatial resolution decreased. While we find that the rSFMS slope at low masses ($\gamma$) is roughly constant with changing spatial resolution, we find that the slope at high masses ($\gamma'$) becomes steeper as spatial resolution decreases. The `spaxel' samples used by \cite{Hani20} appear to be biased toward stellar masses greater than the turnover masses we have identified ($\Sigma_*^{\rm to}$), which means the slopes they obtained would be biased toward the high-mass end of the relation.  Thus, we find that our results are consistent with those of \cite{Hani20}.  
\cite{Pessa2021A&A...650A.134P...PHANGS} fit the rSFMS, resolved molecular gas main sequence, and molecular rKS relation with a single power law and found that the slopes were independent of spatial resolution in a spaxel sample drawn from the Physics at High Angular resolutions in Nearby GalaxieS (PHANGS) survey \citep{PHANGS_release_Leroy2021ApJS..255...19L}. \textnormal{For all three relations, the \textnormal{conditional means reported for the} PHANGS data appear to deviate from an ideal power law}. 

\textnormal{IFS-capable telescopes are increasingly common, opening the door to more spatially resolved data for observed galaxies. Specific proposed and ongoing programs and instruments that will collect IFS data for galaxies include the Galaxy Assemby NIRSpec IFS survey \citep[GA-NIFS;][]{GA-NIFS} using JWST, the first generation Extremely Large Telescope instrument High Angular Resolution Monolithic Optical and Near-Infrared Integral field spectrograph \citep[HARMONI;][]{HARMONI}, the Atacama Large Millimetre Array Large Program [CII] Resolved ISM in Star-forming Galaxies \citep[ALMA-CRISTAL; e.g.,][]{ALMA-CRISTAL}, and the Sloan Digital Sky Survey V Local Volume Mapper survey \citep[LVM;][]{LVM}. The combination of these data sets and a measurement technique that is consistent across varying spatial resolutions would enable studies into the redshift evolution of the rSFMS and the other relations.}
\textnormal{Further, spatially resolved maps of SFR, stellar mass, and gas mass can and often have already been obtained from high-resolution simulations such as FIRE-2, Figuring Out Gas and Galaxies in Enzo \citep[FOGGIE; e.g.,][]{FOGGIE}, and the Assembling Galaxies of Resolved Anatomy project \citep[AGORA;][]{AGORA}. The multitude of possible sources for spatially resolved galaxy data, of which we have only named a selection, emphasizes the need for consistent, objective measurements of scaling relations.}

\textnormal{The \cite{Kennicutt1989ApJ...344..685K} results were based on a sample of $15$ galaxies sampled with a spatial resolution of $1-5$ kpc. With a comparative wealth of data available now and in the coming years, it should be possible to constrain the dependence of $\Sigma_{\rm SFR}$ on $\Sigma_*$ and $\Sigma_{\rm gas}$ and explain deviations from these relations in terms of true stochasticity or other physical parameters such as metallicity, turbulence, and magnetic field strength. However, it is critical that the community come to a consensus on best practices for fitting scaling relationships and describing their scatter. As it is, estimates of the power law slopes in the literature result from a variety of different fitting methods which are often not described with sufficient detail for reproduction. This motivates a consistent reanalysis of existing data with clearly defined, physically-motivated methodology. Such a reanalysis could constrain the influence of resolution, SFR diagnostics and other factors on the resulting relationships. Our results indicate that the ridge line technique we present here is well suited for such a reanalysis but there are further improvements that can be made to improve the statistical rigor. We have published \texttt{ScaleRPy} as an open-source code so that further improvements can be made to the methodology as the community comes to a consensus on statistically rigorous best fitting practices \citep[e.g., following][]{Hogg2010arXiv1008.4686H} and physically-motivated functional forms for the star-forming scaling relationships. Simulations that reproduce spatially resolved galaxies, such as TNG100, are an excellent laboratory for testing fitting techniques because there are often known truth values in the form of the analytical models that describe star formation.}

In conclusion, we find that the \textnormal{ridge line technique offers significant benefits over standard least squares when it comes to fitting scaling relationships.} \textnormal{For the rSFMS, fitting the ridge line with a double power law reveals that the }slope at the low-mass end of the rSFMS is independent of spatial resolution and smoothing scale. This method \textnormal{enhances the comparability of rSFMS measurements because it} does not require pre-selection of star-forming spaxels and parameterizes the location at which the rSFMS slope turns over. \textnormal{For all three spatially resolved scaling relationships we have explored here, the ridge line method results in significantly better fits than least squares.} Thus, \textnormal{the ridge line technique} is a promising method for making \textnormal{objective} measurements of scaling relationships that can be compared across studies. \textnormal{We argue that such objective measurements are highly desirable if the physical processes that govern star formation are to be understood. }The open-source \texttt{ScaleRPy} package facilitates the adoption \textnormal{and improvement} of the method presented here for both global and resolved scaling relations. 


\begin{acknowledgements}
     \textnormal{We thank the anonymous reviewer for their incredibly helpful feedback.} We would like to acknowledge the work and documentation provided by the IllustrisTNG team that has made this paper possible. The IllustrisTNG simulations were undertaken with compute time awarded by the Gauss Centre for Supercomputing (GCS) under GCS Large-Scale Projects GCS-ILLU and GCS-DWAR on the GCS share of the supercomputer Hazel Hen at the High Performance Computing Center Stuttgart (HLRS), as well as on the machines of the Max Planck Computing and Data Facility (MPCDF) in Garching, Germany. The additional computational work done for this paper was performed on the Shared Computing Cluster (SCC) which is administered by Boston University’s Research Computing Services. The SCC is located in the Massachusetts Green High Performance Computing Center. This work was partially supported by NSF grant AST-2009397. B.M. acknowledges support by Northeastern University's Future Faculty Postdoctoral Fellowship Program. 
\end{acknowledgements}

\vspace{5mm}
\software{SciPy \citep{2020SciPy-NMeth};
          ScaleRPy \citep{mcdonough_2025_scalerpy};
          Plotly \citep{plotly}}

\bibliography{draft}{}
\bibliographystyle{aasjournal}



\end{document}